# A Dual-Purpose Deep Learning Model for Auscultated Lung and Tracheal Sound Analysis Based on Mixed Set Training

Fu-Shun Hsu[a,b,c], Shang-Ran Huang[c], Chang-Fu Su[a,d,e], Chien-Wen Huang[f], Yuan-Ren Cheng[c,g,h], Chun-Chieh Chen[f], Chun-Yu Wu[i], Chung-Wei Chen[j], Yen-Chun Lai[c,k], Tang-Wei Cheng[c], Nian-Jhen Lin[c,l], Wan-Ling Tsai[c], Ching-Shiang Lu[c], Chuan Chen[c], and Feipei Lai[a,*]

[a] Graduate Institute of Biomedical Electronics and Bioinformatics, National Taiwan University, Taipei, Taiwan

[b] Cardiovascular Intensive Care Unit, Far Eastern Memorial Hospital, New Taipei, Taiwan

[c] Heroic Faith Medical Science Co., Ltd., Taipei, Taiwan

[d] Department of Anesthesia, Division of Medical Quality, En Chu Kong Hospital, New Taipei, Taiwan

[e] Department of Electronic Engineering, Oriental Institute of Technology, New Taipei, Taiwan

[f] Avalanche Computing Inc., Taipei, Taiwan

[g] Department of Life Science, College of Life Science, National Taiwan University, Taipei, Taiwan

[h] Institute of Biomedical Sciences, Academia Sinica, Taipei, Taiwan

[i] Department of Anesthesiology, National Taiwan University Hospital, Taipei, Taiwan

[j] Department of Surgical Intensive Care Unit, Far Eastern Memorial Hospital, New Taipei City,

Taiwan

k Department of Anesthesiology, Taipei Medical University Hospital, Taipei, Taiwan

l Division of Pulmonary Medicine, Far Eastern Memorial Hospital, New Taipei, Taiwan

***Corresponding Author**: Feipei Lai, Graduate Institute of Biomedical Electronics and Bioinformatics, National Taiwan University, Taipei, Taiwan. (phone: 886-2-3366-4961; Fax: 886-2-3366-3754; e-mail: publicationfpl@gmail.com

# Abstract

Many deep learning-based computerized respiratory sound analysis methods have previously been developed. However, these studies focus on either lung sound only or tracheal sound only. The effectiveness of using a lung sound analysis algorithm on tracheal sound and vice versa has never been investigated. Furthermore, no one knows whether using lung and tracheal sounds together in training a respiratory sound analysis model is beneficial. In this study, we first constructed a tracheal sound database, HF_Tracheal_V1, containing 10448 15-s tracheal sound recordings, 21741 inhalation labels, 15858 exhalation labels, and 6414 continuous adventitious sound (CAS) labels. HF_Tracheal_V1 and our previously built lung sound database, HF_Lung_V2, were either combined (mixed set), used one after the other (domain adaptation), or used alone to train convolutional neural network bidirectional gate recurrent unit models for inhalation, exhalation, and CAS detection in lung and tracheal sounds. The results revealed that the models trained using lung sound alone performed poorly in tracheal sound analysis and vice versa. However, mixed set training or domain adaptation improved the performance for 1) inhalation and exhalation detection in lung sounds and 2) inhalation, exhalation, and CAS detection in tracheal sounds compared to positive controls (the models trained using lung sound alone and used in lung sound analysis and vice versa). In particular, the model trained on the mixed set had great flexibility to serve two purposes, lung and tracheal sound analyses, at the same time.

**Keywords:** Auscultation, convolutional neural network, deep learning, domain adaptation, gated-recurrent unit, lung sound, tracheal sound

# 1. Introduction

Respiratory auscultation [1] with a stethoscope is a longstanding diagnostic technique used to examine the respiratory system. Depending on the auscultation locations, different types of respiratory sounds, such as mouth, tracheal, bronchial, bronchovesicular, and vesicular (lung) sounds, can be heard [2]. Lung and tracheal sounds are the most frequently auscultated in clinical applications.

Lung auscultation is commonly used as a first-line physical examination tool to diagnose pulmonary disease because it is noninvasive and inexpensive [3]. Healthy lungs generate normal lung sounds during breathing; unhealthy lungs may manifest continuous adventitious sounds (CAS) such as wheezes, stridor, or rhonchi or manifest discontinuous adventitious sounds (DAS) such as crackles or pleural friction rubs [1, 2]. Health care professionals can recognize abnormal pulmonary conditions by the presence, type, characteristics, and location of adventitious lung sounds [1-3].

Tracheal auscultation [3] over the pre-tracheal or pre-cordial region can be used to detect pulmonary ventilation abnormalities such as abnormal respiratory rates, upper airway obstructions, and apnea. Respiratory rates can be estimated by identifying breath phases (inhalation and exhalation) in the tracheal sound [4, 5]. Partial upper airway obstruction is indicated by the presence of CAS-like patterns (green arrows in Fig. 1) such as stridor [6, 7] or snoring [8]. Total upper airway obstruction is indicated by the pattern of a very short and high-intensity sound (white arrows in Fig. 1) which represents a prematurely stopped inhalation resulting from high airway resistance and airway collapse. Apnea can be inferred from the prolonged absence of inhalation and exhalation during tracheal auscultation [5, 7, 9-11]. Therefore, tracheal auscultation is recommended by some clinical guidelines for use if a patient's pulmonary ventilatory functions are at risk of being compromised, such as in monitored anesthesia care [12-15].

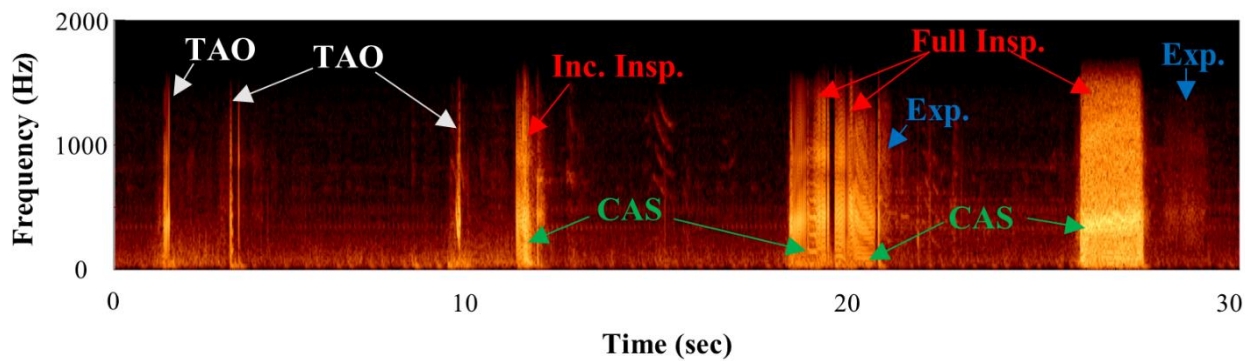

**Fig 1. The patterns of total airway obstruction (white arrows), an incomplete and full inspirations (red arrows), expirations (blue arrows), and CAS (green arrows) displayed on a spectrogram of 30-second tracheal sound.** TAO: total airway obstruction. Inc.: incomplete, Insp.: inspiration. Exp.: expiration. CAS: continuous adventitious sound.

Although respiratory auscultation using a conventional stethoscope can be used in many clinical applications, it is gradually forgotten in daily practice because of some limitations [16, 17] and the advent of advanced diagnostic devices. Also, sudden loud sounds, such as a patient's talking and unconscious mumbling, and drilling and spurting sounds in a dental procedure, may hurt the practitioner's ears during conventional auscultation. Computerized respiratory sound analysis [18, 19] is key to overcoming the limitations of the old auscultation technique, which can help health care professionals hear and analyze the auscultated sound. Previously proposed machine learning-based [2, 20] and deep learning-based algorithms [17, 21] in computerized respiratory sound analysis have been comprehensively reviewed. However, no one has investigated whether a lung sound analysis model can make an accurate prediction in tracheal sounds and vice versa. Moreover, a larger size of training data set is commonly thought to be an important factor in training a better deep learning model [22, 23]. However, no one knows when the lung and tracheal sound data are combined to form a larger training data set whether the trained model can perform better in the lung and tracheal sound analyses.

To investigate the above problems, we need large amounts of accessible lung and tracheal sound data. In our previous studies, we had established the lung sound databases HF_Lung_V1 [16] and HF_Lung_V2 [24]. Deep learning–based convolutional neural network (CNN)-bidirectional gated recurrent unit (BiGRU) models were trained and demonstrated to adequately detect inhalation, exhalation, CAS, and DAS events in lung sounds at a recording level [16, 24]. However, so far there is no large open-access tracheal sound data set. Thus, in this study, we first established a tracheal sound database, HF_Tracheal_V1 (Access to audio files can be found at https://gitlab.com/techsupportHF/HF_Tracheal_V1. Access to label files must be granted by filling out an application form at https://forms.gle/CDew586M1x4yQprw7. The addresses will be changed and fixed after the manuscript revision). Then, we trained a deep learning-based lung sound model using lung sound data and tested how good the model was for detecting inhalation, exhalation, and CAS in tracheal sounds, and vice versa. Further, to investigate the feasibility and benefit of using lung and tracheal sounds together in training lung and tracheal sound analysis models, two feasible approaches were identified. First, the lung and tracheal sound files could be combined to form a mixed set to train a single model for both lung and tracheal sound analysis. Second, transfer learning [25], specifically domain adaptation [26], could be used to fine-tune a pretrained lung sound model for tracheal sound analysis (or vice versa) to further improve the model performance. In addition, we also investigated how well a single model can serve dual purposes, lung and tracheal sound analysis, at the same time.

## 2. Related Work

Computerized respiratory sound analysis methods have been thoroughly reviewed [2, 20]. However, most previously reported methods were machine learning-based and limited by a small data set. Not until recently have many deep learning-based analysis methods been proposed [17, 21, 27]. Nevertheless, most of the studies focused only on lung sound classification tasks, such as classifying subjects into healthy ones or the ones with pulmonary diseases [28-31] and classifying respiratory sound into normal one or the one containing a certain type of adventitious sound [32-35]. Only few studies investigated event detection at a recording level [36, 37]. By contrast, research on tracheal sound analysis using deep learning [38, 39] is uncommon. No study has explored how much difference in the acoustic features of lung and tracheal sound can influence training a computerized respiratory sound analysis model, and no one has investigated the feasibility of combining lung and tracheal sound data sets to train a better model.

## 3. Materials and Methods

### 3.1 Participants

The tracheal sounds were obtained in several studies. The protocol for the tracheal sound recording was reviewed and approved by the following Institutional Review Boards (IRBs), including the Joint Institutional Review Board of the Medical Research Ethical Foundation, Taipei, Taiwan (case number: 19-006-A-2), IRB of Taipei Tzu Chi Hospital, Buddhist Tzu Chi Medical Foundation (case number: 09-XD-079), IRB of En Chu Kong Hospital (case number: ECKIRB1090303), IRB of Taipei City Hospital (case number: TCHIRB-10811009-E), and IRB of National Taiwan University Hospital (case numbers: 201811991DIND and 202005108DIPC). This study was conducted in accordance with the 1964 Helsinki Declaration and its later amendments.

A total of 227 Taiwanese individuals (110 men, 116 women, and 1 unknown) aged ≥20 years who had undergone a diagnostic or surgical procedure, including laparoscopy (n = 1), gastrointestinal endoscopy (n = 209), cesarean section (n = 8), and obstructive sleep apnea-related procedures (n = 9), under monitored anesthesia care were enrolled in this study. The average age was 51.3 ± 13.3 years. The average height and weight were 164.6 ± 8.6 cm and 66.0 ± 13.4 kg. The average body mass index was 24.2 ± 3.6 kg/m$^2$. The demographic data of the participants are summarized in Table 1. We did not enroll individuals belonging to vulnerable groups (e.g., incarcerated individuals, indigenous people, persons with disabilities, or persons with mental illness), having a history of allergies that prevented contact with medical patches or artificial skin, or having a diagnosis of atrial fibrillation or arrhythmia. Tracheal sounds were collected between November 2019 and June 2020.

**Table 1. Demographic characteristics of participants.**

| | |
|---|---|
| **Number (N)** | 227 |
| **Sex (M/F/unknown)** | 110/116/1 |
| **Age (year)** | 51.3 ± 13.3* |
| **Height (cm)** | 164.6 ± 8.6† |
| **Weight (kg)** | 66.0 ± 13.4† |
| **BMI (kg/m$^2$)** | 24.2 ± 3.6† |
| **Procedure** | |
| **Laparoscopy** | 1 (0.4%) |
| **GI Endoscopy** | |
| UGI | 2 (0.9%) |
| LGI | 33 (14.5%) |
| UGI + LGI | 108 (47.6%) |
| Unspecified | 66 (29.1%) |
| **Cesarean Section** | |
| Extraperitoneal | 5 (2.2%) |
| Unspecified | 3 (1.3%) |
| **Obstructive Sleep Apnea-Related** | |
| DISE | 6 (2.6%) |
| DISE+SMP | 1 (0.4%) |
| UPPP+SMP | 1 (0.4%) |
| UPPP+SMP+DISE | 1 (0.4%) |

GI: Gastrointestinal, UGI: upper GI, LGI: lower GI, DISE: drug-induced sleep endoscopy, SMP: septomeatoplasty, and UPPP: uvulopalatopharyngoplasty. *: 7 missing data. †: 9 missing data.

## 3.2 Tracheal sound recording

Two devices, HF-Type-2 and HF-Type-3 devices, were used to record tracheal sounds. HF-Type-

2 (Fig. 2a) is an electronic stethoscope (AS-101, Heroic Faith Medical Science, Taipei, Taiwan) connected to a smartphone (Mi 9T Pro, Xiaomi, Beijing, China). HF-Type-3 (Fig. 2b) is constructed from a chestpiece (603P, Spirit Medical, New Taipei, Taiwan), a stethoscope tubing, a microphone (ECM-PC60, Sony, Tokyo, Japan), and a smartphone (Mi 9T pro, Xiaomi, Beijing, China). A customized app was installed in the smartphone to record the received tracheal sounds. Tracheal sounds from each participant were recorded at the flat area of the left or right thyroid cartilage as displayed in Fig. 3 by using one of the devices. Although HF-Type-2 supported multichannel recording, only one channel was used for tracheal sound recording. Tracheal sounds were collected at a sampling rate of 4000 Hz and 16-bit depth. Tracheal sounds were recorded while the participants were undergoing a procedure under monitored anesthesia care. The recording began before the first administration of the anesthetic and ended when the procedure was completed (before they were transferred to a post-anesthesia care unit). The recording time varied depending on the necessity of tracheal sound monitoring; most recordings ranged from a few minutes to less than 20 minutes. The majority of the recordings were collected when the participants were under moderate sedation; however, some recording covered the periods of mild and deep sedation [40]. The numbers of patients whose tracheal sound was recorded by HF-Type-2 and HF-Type-3 were 108 and 119, respectively.

### 3.3 Data labeling

The collected tracheal sound recordings were first truncated to 15-s files using a sliding window with a step size of 15 s; thus, the truncated files did not overlap. Any tracheal sound file shorter than 15 s was deleted.

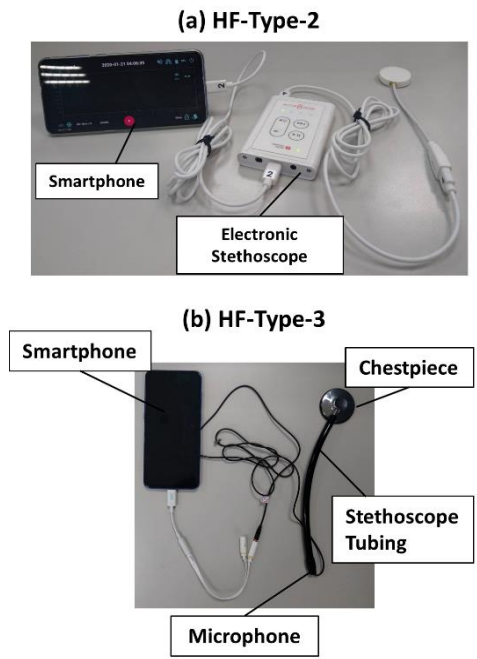

Fig. 2. (a) HF-Type-2 and (b) HF-Type-3 devices used for tracheal sound recording.

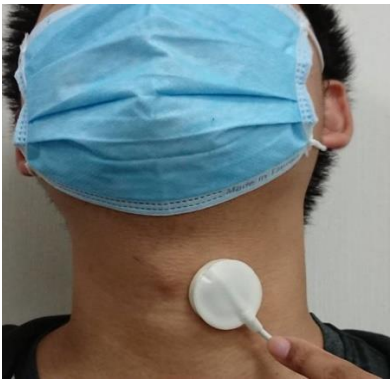

Fig. 3. Tracheal sound recording location.

Each of the 15-s audio files was subsequently labeled by one labeler and inspected by an inspector.

The labeling was done either by a respiratory therapist (NJL) with 8 years of clinical experience or a nurse (WLT) with 13 years of clinical experience. The quality of the labeling was verified by another respiratory therapist (CC) with 6 years of clinical experience or another nurse (CSL) with 4 years of clinical experience. The labelers' and inspectors' occupational licenses were granted by the Examination Yuan, Taiwan, and they practiced their profession in Taiwan.

A self-developed labeling software [41] was used for labeling. If the inspector and the labeler did not agree on a label, it was further reviewed and amended until mutual agreement was reached. The consensus on labeling criteria were maintained throughout the labeling process by holding regular meetings. Labelers were asked to label the start and end times of inhalation (I), exhalation (E), and CAS (C) events. Unlike labels in HF_Lung_V1 and HF_Lung_V2, we did not specifically differentiate tracheal sound CAS events as a wheeze, stridor, or rhonchus. However, CAS labels in this study included snoring. We did not label DAS in this study because crackles and pleural friction rubs were rarely present in the tracheal sounds.

The tracheal sound recordings and the corresponding labels were divided into a training set and a testing set at a ratio of approximately 4:1. Files from the same participant were all assigned to the same set (either training or testing). The training set and testing set formed the HF_Tracheal_V1 (Tracheal_V1) database.

## 3.4 Deep learning pipeline

The CNN-BiGRU model (Fig. 4) outperformed all the other benchmark models in the lung sound analysis of our previous study [16]. Therefore, the CNN-BiGRU model was used in this study.

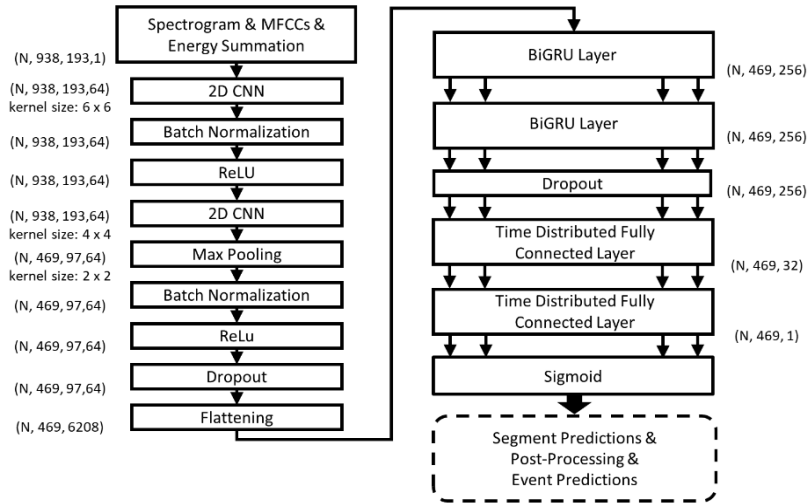

**Fig. 4. Architecture of the CNN-BiGRU model.** This figure is adapted from our previous study [42] under the Creative Commons Attribution (CC BY) license.

The major tasks for the models were to do inhalation, exhalation and CAS event detection at the recording level, which was clearly defined in our previous studies [16, 24]. The deep learning pipeline is presented in Fig. 5, and it was the same as that in our previous studies [16, 24]. The 15-s signals were first filtered by a Butterworth high-pass filter with a cutoff frequency at 80 Hz. The spectrogram was then computed from the 15-s filtered signal using a short-time Fourier transform (STFT) [43] with a Hanning window of size 256, a hop length of size 64, and no zero padding. A 938 × 129 matrix was generated; 938 was the number of time frames (segments) and 129 was the number of frequency bins. The mel frequency cepstral coefficients (MFCCs) [37], including 20 static coefficients, 20 delta coefficients, and 20 acceleration coefficients were derived from every time segment of the spectrogram resulting in three 938 × 20 MFCC matrices. The energy in the four frequency bands of the spectrogram (0–250, 250–500, 500–1000, and 0–2000 Hz) was summed to produce four 938 × 1 energy summation vectors. The spectrogram, each of the three MFCC matrices, and each of the energy summation vectors were then normalized. The concatenation of the normalized spectrogram, MFCCs, and energy summation were fed into the CNN-BiGRU model. The output of the CNN-BiGRU model was a 469 × 1 probability vector. Thresholding was then applied to the probability vector to generate a binarized vector. Elements of the binary vector with a value of 1 indicated that sounds of inhalation, exhalation, or CAS were detected in the corresponding time segment. After the segment detection results were obtained, the vectors were sent for postprocessing to merge neighboring connected segments into events. Then, we examined if any two events were spanned only by a short interval (<0.5 sec) and their maximum energy peaks were in a very close frequency range (Δ25 Hz), we chose to merge these events. Finally, we removed burst events (duration <0.05 sec) to generate the event-detection results.

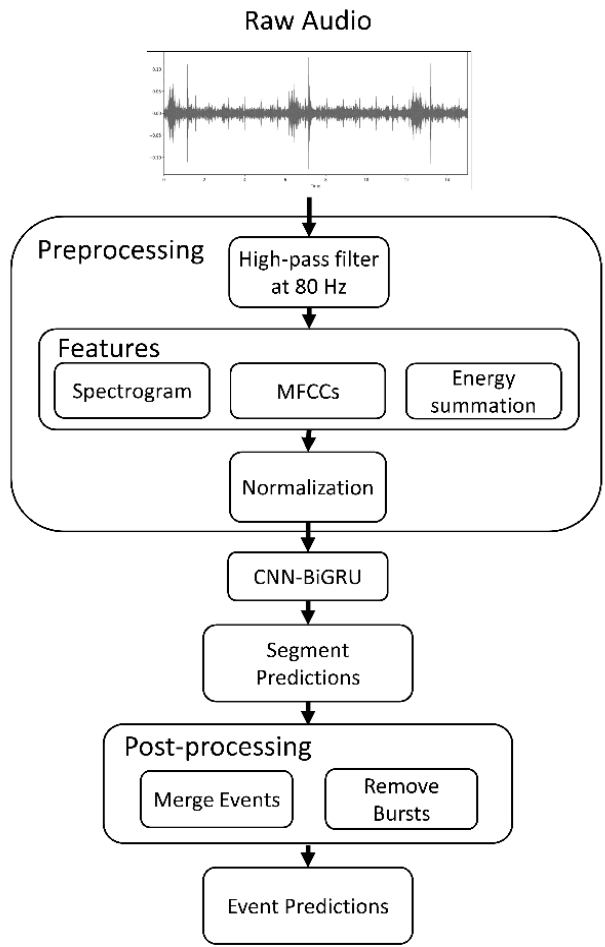

**Fig. 5. Deep learning pipeline.** This figure is adapted from our previous study [42] under the Creative Commons Attribution (CC BY) license.

## 3.5 Investigation of mixed set training and domain adaptation

To investigate whether using the lung and tracheal sound data together can train a better model for lung and tracheal sound analysis, different training strategies were applied. First, full training [44] (training from scratch) was used. Lung sound models were trained using the training set of HF_Lung_V2 (Lung_V2_Train) alone, and tracheal sound models were trained using the training set of HF_Tracheal_V1 (Tracheal_V1_Train) alone. Second, the recordings in Lung_V2_Train and Tracheal_V1_Train were combined, and models were trained on this mixed set. Third, we used domain adaptation [26] to fine-tune the pretrained lung sound models for tracheal sound analysis, and we fine-tuned the pretrained tracheal sound models for lung sound analysis. All trained models were tested separately on both testing sets of HF_Lung_V2 (Lung_V2_Test) and testing set of HF_Tracheal_V1 (Tracheal_V1_Test). The positive controls (PCs) were 1) models trained on Lung_V2_Train alone and tested on Lung_V2_Test and 2) models trained on Tracheal_V1_Train alone and tested on Tracheal_V1_Test. The negative controls (NCs) were 1) models trained on Lung_V2_Train alone and tested on Tracheal_V1_Test and 2) models trained on Tracheal_V1_Train alone and tested with Lung_V2_Test. Only recordings containing at least one I, E or C label were used to train the corresponding detection model and evaluate the model performance.

### 3.6 Training environment and parameters

Models were trained on a server (OS: Ubuntu 18.04; CPU: Intel Xeon Gold 6154@3.00 GHz; RAM: 90 GB) provided by the National Center for High-performance Computing in Taiwan. We used TensorFlow 2.10 as the framework for the deep neural networks. GPU acceleration was provided by an NVIDIA Titan V100 GPU on the CUDA 10 and CuDNN 7 frameworks.

To train the model, we set the batch size to 64 and the number of epochs to 5000. Note that we also set the early stop policy with a patience of 50 during the training procedure. We used Adam optimizer with an initial learning rate of 0.0001 for training.

For domain adaptation, we used the model trained on either HF_Lung_V2 or HF_Tracheal_V1

as the pretrained model without freezing the weights. Then, we fine-tune the pre-trained model for 50 epochs with the training parameters mentioned above.

In this study, fivefold cross validation was applied and repeated thrice. Hence, eventually, we had 15 trained models for each training scenario. Note that only audio files containing I, E, and C labels were used to train and test the corresponding models.

### 3.7 Performance evaluation

The segment-detection and event-detection performance of the models at the recording level was evaluated [16, 42]. We first had the ground-truth event labels in the 15-s recordings (red horizontal bars in Fig. 6a) after the labelers did the labeling. After we extracted the features in the deep learning pipeline, a 15-s recording was broken down into 938 time frames (segments). Then, we assigned a segment as a ground-truth segment (red vertical bars in Fig. 6b) if more than half of its width overlapped with a ground-truth event. The results of segment prediction (blue vertical bars in Fig. 6c) were obtained after the model finished the inference process. After the segment prediction results postprocessing, the event-prediction results were obtained (blue horizontal bars in Fig. 6d). With the ground-truth labels and prediction results, we can evaluate whether a model made a correct detection. Each predicted segment was compared with the ground-truth segment; a perfectly matched segment was a true positive (TP) segment (orange vertical bars in Fig. 6e); if the model falsely predicted a segment containing a target event, the segment was a false positive (FP) event (black vertical bars in Fig. 6e); if the model failed to detect a target event in a segment, the segment was a false negative (FN) segment (yellow vertical bars in Fig. 6e); if the model agreed with the ground truth that a segment did not contain a target event, the segment was assigned as a true negative (TN) segment (green vertical bars in Fig. 6e) Similarly, we used the Jaccard index (JI) [36], described in equation (1), to determine whether the models correctly detect an event.

$$Jaccard\ index\ (JI) = \frac{period\ of\ a\ target\ event\ \cap\ period\ of\ a\ referenced\ event}{period\ of\ a\ traget\ event\ \cup\ period\ of\ a\ referenced\ event} \qquad (1)$$

First, we used the ground-truth labels as a reference and examined whether each ground-truth label had a matching predicted event (JI ≥ 0.5). If the label had such an event, it was counted as a TP event (orange horizontal bar in Fig. 6f); otherwise, it was counted as an FN event (yellow horizontal bars in Fig. 6f). Then, conversely, we used the event prediction results as a reference; we checked whether we could find a matching ground-truth label for each predicted event (JI ≥ 0.5). If we could do so, the predicted event was designated as a TP event (orange horizontal bar in Fig. 6g); if not, it was counted as an FP event (black horizontal bars in Fig. 6g). TN events were not counted because the background phase was not considered an event. Note that TP events were counted twice (orange horizontal bars in Fig. 6f and Fig. 6g) during the evaluation process. Therefore, a pair of TPs was considered a single TP event in computing the evaluation metrics. However, all the FP and FN events were used to compute the evaluation metrics despite the possibility of inducing undesirable bias.

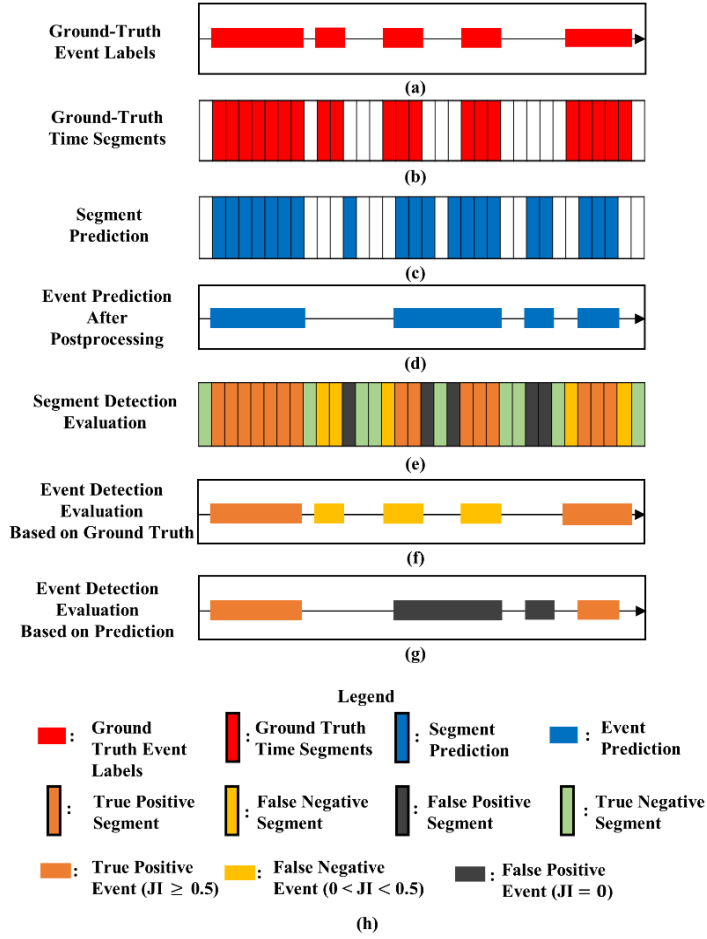

**Fig 6. Illustration of event detection evaluation.** (a) Ground-truth event labels, (b) segment prediction, (c) event prediction after postprocessing, (d) event detection evaluation based on ground-truth event labels, (e) event detection evaluation based on event prediction, and (f) legend. JI: Jaccard index. This figure is reproduced from our previous study [42] under the Creative Commons Attribution (CC BY) license.

The following indexes were used for performance evaluation:

$$\text{accuracy} = (TP + TN) / (TP + TN + FP + FN), \quad (2)$$

$$\text{sensitivity} = TP / (TP + FN), \quad (3)$$

$$\text{specificity} = (TN) / (TN + FP), \quad (4)$$

$$\text{positive predictive value (PPV)} = TP / (TP + FP), \quad (5)$$

$$\text{F1 score} = 2 \times (\text{sensitivity} \times \text{PPV}) / (\text{sensitivity} + \text{PPV}). \tag{6}$$

The threshold producing the best accuracy for segment prediction was used to generate event-prediction results. We calculated accuracy, sensitivity, specificity, PPV, F1 score [37], and area under the curve of receiver operating characteristic (AUC) to evaluate the segment-detection performance. We used sensitivity, PPV, and F1 score to evaluate the event-detection performance; accuracy, specificity, and AUC were not used for performance evaluation of the event-detection because a TN event was not defined. In this paper, we focused on reporting the event-detection performance and moved the results of segment detection to the supplementary because event detection was our end goal. Because we had 15 models (3 repetitions of fivefold cross validation) in each training scenario, the mean value ± standard deviation of the indexes was reported.

To evaluate a model's potential of being used in lung and tracheal sound analyses at the same time, we took the mean of the two values of an event-detection performance index derived from the respective Lung_V2_Test and Tracheal_V1_Test weighted by the numbers of labels.

## 3.8 Statistical comparison

The mean duration of I, E, and C labels between datasets were compared using Student's t-test. The comparison of the performance indexes between different training scenarios was conducted using Wilcoxon rank sum test. The statistical comparison was conducted using MATLAB R2022a (MathWorks Inc., Natick, Massachusetts, USA).

# 4. Results

## 4.1 Summary of HF_Lung_V2 and HF_Tracheal_V1 databases

The summary statistics of the HF_Lung_V2 and HF_Tracheal_V1 databases are listed in Table 2. A total of 13957 15-s recordings, 49373 I labels, 24552 E labels, and 21558 C labels were included in HF_Lung_V2. Tracheal_V1 contains 10448 15-s recordings, 21741 I labels, 15858 E labels, and 6414 C labels. The mean duration of I labels significantly differed ($p < 0.001$) between HF_Lung_V2 and HF_Tracheal_V1 (0.95 ± 0.30 and 1.10 ± 0.34 s, respectively). The mean durations of E labels significantly differed ($p < 0.001$) between HF_Lung_V2 and HF_Tracheal_V1 (0.91 ± 0.49 and 1.00 ± 0.40 s, respectively). The mean durations of C labels significantly differed between HF_Lung_V2 and HF_Tracheal _V1 (0.82 ± 0.46 and 1.08 ± 0.58 s, respectively).

**Table 2. Summary of HF_Lung_V2 and HF_Tracheal_V1 databases.**

| Recordings or Labels | Attributes | HF_Lung_V2 | HF_Tracheal_V1 |
|---|---|---|---|
| Recordings | No. of 15 s recordings | 13957 | 10448 |
|  | Total duration (min) | 3489.25 | 2739.5 |
| I | No. | 49373 | 21741 |
|  | Total duration (min) | 785.48 | 397.96 |
|  | Mean duration (s) | 0.95 ± 0.30 | 1.10 ± 0.34* |
| E | No. | 24552 | 15858 |
|  | Total duration (min) | 374.24 | 263.08 |
|  | Mean duration (s) | 0.91 ± 0.49 | 1.00 ± 0.40* |
| C | No. | 21558 | 6414 |
|  | Total duration (min) | 292.85 | 115.06 |
|  | Mean duration (s) | 0.82 ± 0.46 | 1.08 ± 0.58* |

I: inhalation labels, E: exhalation labels, and C: continuous adventitious sound labels.

Mean duration is presented as mean ± standard deviation. *$p < 0.001$ between HF_Lung_V2 and HF_Tracheal_V1.

## 4.2 Statistics of training and test data sets

The compositions of the training and testing data sets of both the HF_Lung_V2 and HF_Tracheal_V1 databases are summarized in Table 3. Lung_V2_Train and Tracheal_V1_Train had 10554 and 8398 15-s files, respectively, and Lung_V2_Test and Tracheal_V1_Test had 3403 and 2050 15-s files, respectively. The statistics of the I, E, and C labels in the training and test data sets of HF_Lung_V2 and HF_Tracheal_V1 are presented in Table 3. The mean durations of I, E, and C labels between all pairs of Lung_V2_Train, Tracheal_V1_Train, Lung_V2_Test, and Tracheal_V1_Test significantly differed ($p < 0.001$ for all).

**Table 3. Training and test data sets of HF_Lung_V2 and HF_Tracheal_V1.**

| Recordings or Labels | Attributes | Training datasets | | Test datasets | | p-value |
|---|---|---|---|---|---|---|
| | | Lung_V2 | Tracheal_V1 | Lung_V2 | Tracheal_V1 | |
| Recordings | No. of 15 s recordings | 10554 | 8398 | 3403 | 2050 | |
| | Total duration (min) | 2638.5 | 2099.5 | 850.75 | 512.5 | |
| I | No. | 39057 | 17957 | 10316 | 3784 | |
| | Total duration (min) | 623.02 | 333.86 | 162.46 | 64.09 | |
| | Mean duration (s) | 0.96 ± 0.30 | 1.12 ± 0.33 | 0.94 ± 0.26 | 1.02 ± 0.34 | <0.001* |
| E | No. | 18334 | 13162 | 6218 | 2696 | |
| | Total duration (min) | 292.88 | 221.30 | 81.37 | 41.77 | |
| | Mean duration (s) | 0.96 ± 0.52 | 1.01 ± 0.40 | 0.79 ± 0.37 | 0.93 ± 0.36 | <0.001* |
| C | No. | 17361 | 5871 | 4197 | 543 | |
| | Total duration (min) | 240.41 | 106.06 | 52.45 | 9.00 | |
| | Mean duration (s) | 0.83 ± 0.48 | 1.08 ± 0.57 | 0.75 ± 0.34 | 1.00 ± 0.68 | <0.001* |

I: inhalation labels, E: exhalation labels, and C: continuous adventitious sound labels. * indicates that mean durations of I, E, and C labels between all pairs of Lung_V2_Train, Tracheal_V1_Train, Lung_V2_Test, and Tracheal_V1_Test significantly differed ($p < 0.001$).

## 4.3 Model performance

The event-detection performance of the trained models are presented in Table 4. Apparently, the NCs had the worst performance compared with the other models; all performance indexes were significantly worse than the ones of the PCs by a large margin ($p < 0.001$). The models trained on a mixed set had the best performance in all three indexes for inhalation and exhalation detection in lung sound analysis; however, the models first trained on Lung_V2_Train and then finetuned with Tracheal_V1_Train (Lung_V2 → Tracheal_V1) had the best performance for inhalation and exhalation detection in tracheal sound analysis. For CAS detection, the PC slightly won over the one trained on the basis of domain adaption in the lung sound analysis, and the Lung_V2→Tracheal_V1 CAS model displayed best performance in tracheal sound analysis.

**Table 4. Performance of event detection of the trained models.**

| Controls | Training Database/Strategy | Testing Database | Event Detection PPV | SEN | F1-Score |
|---|---|---|---|---|---|
| Inhalation | | | | | |
| PC | Lung_V2 | Lung_V2 | 0.832±0.010 | 0.869±0.011 | 0.850±0.009 |
| NC | Tracheal_V1 | Lung_V2 | 0.736±0.015††† | 0.684±0.022††† | 0.709±0.018††† |
| | Lung_V2+Tracheal_V1 | Lung_V2 | **0.839±0.007** | **0.872±0.009** | **0.855±0.006** |
| | Tracheal_V1→Lung_V2 | Lung_V2 | 0.829±0.012 | 0.870±0.009 | 0.849±0.010 |
| NC | Lung_V2 | Tracheal_V1 | 0.610±0.021††† | 0.620±0.029††† | 0.615±0.024††† |
| PC | Tracheal_V1 | Tracheal_V1 | 0.793±0.011 | 0.801±0.017 | 0.797±0.014 |
| | Lung_V2+Tracheal_V1 | Tracheal_V1 | 0.782±0.013 | 0.804±0.016 | 0.793±0.013 |
| | Lung_V2→Tracheal_V1 | Tracheal_V1 | **0.802±0.011** | **0.822±0.018**** | **0.812±0.014**** |
| Exhalation | | | | | |
| PC | Lung_V2 | Lung_V2 | 0.754±0.017 | 0.749±0.019 | 0.751±0.017 |
| NC | Tracheal_V1 | Lung_V2 | 0.491±0.032††† | 0.412±0.032††† | 0.448±0.031††† |
| | Lung_V2+Tracheal_V1 | Lung_V2 | **0.788±0.009**\*\*\* | **0.776±0.018**\*\*\* | **0.782±0.013**\*\*\* |
| | Tracheal_V1→Lung_V2 | Lung_V2 | 0.771±0.010\*\* | 0.764±0.012\* | 0.767±0.010\*\* |
| NC | Lung_V2 | Tracheal_V1 | 0.501±0.048††† | 0.444±0.060††† | 0.471±0.054††† |
| PC | Tracheal_V1 | Tracheal_V1 | 0.789±0.007 | 0.802±0.008 | 0.796±0.007 |
| | Lung_V2+Tracheal_V1 | Tracheal_V1 | 0.788±0.011 | 0.806±0.009 | 0.797±0.009 |
| | Lung_V2→Tracheal_V1 | Tracheal_V1 | **0.796±0.009**\* | **0.813±0.009**\*\* | **0.805±0.008**\*\* |
| Continuous adventitious sound | | | | | |
| PC | Lung_V2 | Lung_V2 | **0.438±0.023** | 0.403±0.028 | **0.420±0.025** |
| NC | Tracheal_V1 | Lung_V2 | 0.315±0.022††† | 0.315±0.027††† | 0.315±0.024††† |
| | Lung_V2+Tracheal_V1 | Lung_V2 | 0.411±0.025†† | 0.377±0.031† | 0.393±0.028†† |
| | Tracheal_V1→Lung_V2 | Lung_V2 | 0.431±0.031 | **0.406±0.039** | 0.418±0.036 |
| NC | Lung_V2 | Tracheal_V1 | 0.441±0.080††† | 0.431±0.079††† | 0.436±0.079††† |
| PC | Tracheal_V1 | Tracheal_V1 | 0.661±0.047 | 0.680±0.054 | 0.670±0.050 |
| | Lung_V2+Tracheal_V1 | Tracheal_V1 | **0.780±0.024**\*\*\* | 0.794±0.032\*\*\* | 0.787±0.027\*\*\* |
| | Lung_V2→Tracheal_V1 | Tracheal_V1 | 0.777±0.021\*\*\* | **0.807±0.023**\*\*\* | **0.792±0.021**\*\*\* |

PC: positive control, and NC: negative control. Bold values indicate better performance. Lung_V2+Tracheal_V1 denotes mixed set training. Tracheal_V1→Lung_V2 denotes pretrained tracheal sound model that was fine-tuned with the lung sound data. Lung_V2→Tracheal_V1 denotes the pretrained lung sound model that was fine-tuned with the tracheal sound data. All values are presented as mean ± standard deviation.

†, ††, and ††† represent significantly worse performance (p <0.05, p <0.01, and p <0.001, respectively) compared with PC.

\*, \*\*, and \*\*\* represent significantly better performance (p <0.05, p <0.01, and p <0.001, respectively) compared with PC.

Table 5 displays the weighted mean values of event-detection indexes derived by averaging the two scores in the Lung_V2_Test and Tracheal_V1_Test. The results clearly demonstrate that models trained with a mixed set had the best performance when they were used in lung and tracheal sound analysis at the same time.

**Table 5. Adjusted mean values of event detection performance indexes for inhalation, exhalation, and CAS detection derived by averaging the two scores in the Lung_V2_Test and Tracheal_V1_Test weighted by the numbers of labels.**

| Tasks | Training Database/Strategy | Event Detection | | |
|---|---|---|---|---|
| | | PPV | SEN | F1-Score |
| Inhalation | Lung_V2 | 0.773 | 0.802 | 0.787 |
| | Tracheal_V1 | 0.751 | 0.715 | 0.732 |
| | Lung_V2+Tracheal_V1 | **0.824** | **0.854** | **0.838** |
| | Tracheal_V1→Lung_V2 | 0.775 | 0.811 | 0.793 |
| | Lung_V2→Tracheal_V1 | 0.789 | 0.772 | 0.780 |
| Exhalation | Lung_V2 | 0.678 | 0.657 | 0.666 |
| | Tracheal_V1 | 0.581 | 0.530 | 0.553 |
| | Lung_V2+Tracheal_V1 | **0.788** | **0.785** | **0.786** |
| | Tracheal_V1→Lung_V2 | 0.734 | 0.722 | 0.728 |
| | Lung_V2→Tracheal_V1 | 0.694 | 0.644 | 0.667 |
| CAS | Lung_V2 | 0.438 | 0.407 | 0.422 |
| | Tracheal_V1 | 0.355 | 0.357 | 0.355 |
| | Lung_V2+Tracheal_V1 | **0.453** | **0.425** | **0.438** |
| | Tracheal_V1→Lung_V2 | 0.434 | 0.411 | 0.422 |
| | Lung_V2→Tracheal_V1 | 0.409 | 0.403 | 0.403 |

CAS: continuous adventitious sound. Bold values indicate the best performance among the five models. Lung_V2+Tracheal_V1 denotes mixed set training. Tracheal_V1→Lung_V2 denotes the pretrained tracheal sound model that was fine-tuned with the lung sound data. Lung_V2→Tracheal_V1 denotes the pretrained lung sound model that was fine-tuned with the tracheal sound data.

## 5. Discussion

Our results reveal that NCs performed the worst compared with the other models. It is because lung and tracheal sounds have different frequency ranges, energy drops, inhalation–exhalation duration ratios, and pause periods [2]. Besides, the mean durations for the I, E, and C labels significantly differed between HF_Lung_V2 and HF_Tracheal_V1 (see Table 2). The majority of feature distribution differences resulted from innate differences in the physical and physiological mechanisms underlying the generation of lung and tracheal sounds [45]. Thus, future studies must consider lung and tracheal sound as two distinct domains when building computerized analysis models. However, it is undeniable that the different recording devices used to record the lung [16, 42] and tracheal sounds may also generate some feature differences. In addition, the setting of the patients, such as use of mechanical ventilation, depth of sedation, body position, etc., may contribute to the divergence of feature distribution.

As presented in Table 4, mixed set training and domain adaptation can improve the performance of inhalation and exhalation detection in lung sound analysis and inhalation, exhalation, and CAS detection in the tracheal sound analysis. The extent of benefits may hinge on the composition of the lung and tracheal sound data; moreover, we can't assure the benefit can still be present when the data sizes of the lung and tracheal sounds grow to a larger scale. Nevertheless, as clearly indicated in Table 5, the model that was trained on a mixed set was capable of processing a testing set comprising mixed lung and tracheal sound data. Mixed set training is an attractive option for developing an all-purpose respiratory monitor; users are not required to pick a specific channel or switch to a specific algorithm for lung or tracheal sound analysis.

The model performance for CAS detection was significantly better for tracheal sounds than for lung sounds (Table 4). This result may be because CAS in tracheal sounds is louder, increasing the signal-to-noise ratio and facilitating CAS pattern identification. Additionally, the C labels in HF_Lung_V2 could be noisy and are currently undergoing reworking [16, 42]. Furthermore, CAS in Tracheal_V1 is thought to be a primarily monophonic event occurring in the inspiratory phase, characterized by extrathoracic upper airway obstructions [46] induced by anesthetic drugs. However, CAS in HF_Lung_V2 can be categorized as inspiratory, expiratory, and biphasic types and as monophonic and polyphonic events [2]. Thus, the CAS features in Tracheal_V1 are not as diverse as those in HF_Lung_V2.

In contrast to the labeling in HF_Lung_V2, DAS was not specifically labeled in HF_Tracheal_V1 because most diseases generating DAS (fine crackles, coarse crackles, and pleural friction rubs) do not occur in the upper airway close to the pretracheal region. However, DAS-like patterns were occasionally observed in the collected tracheal sounds. These patterns might be caused by air flowing through accumulations of fluids such as water, saliva, sputum, or blood in the upper airway. Fluid accumulation in the upper airway must be promptly managed by clinicians, such as those executing a dental procedure on a moderately or deeply sedated patient who is not able to voluntarily cough to expel fluids in the laryngeal region [47]; in such a case, the dental team must perform suction to prevent aspiration. Hence, a respiratory monitor capable of detecting fluid accumulation in the upper airway is of clinical importance. The labeling of DAS-like patterns in tracheal sounds worth consideration in the future research.

In clinical practice, capnography is more commonly used for pulmonary ventilation monitoring than tracheal sound auscultation. Moreover, a pulse oximeter is now required for blood oxygen monitoring in surgical procedures or in a diagnostic procedure involving anesthesia. However, both of these devices have some limitations. Capnographic accuracy can be compromised by the poor sampling of carbon dioxide due to open-mouth breathing [48, 49]; by the use of a face mask or nasal cannula [50-52]; or by procedures that cause airflow interference, such as esophagogastroduodenoscopy or bronchoscopy. Moreover, capnography is difficult to use in surgeries involving the facial or oral regions. Oxygen desaturation as measured by a pulse oximeter is a delayed response to abnormal pulmonary ventilation [53, 54]. Therefore, a tracheal sound monitor that automatically detects abnormal respiratory rates, upper airway obstructions, and apnea could have substantial clinical value and could complement capnography and oximetry [4, 7]. Therefore, more accurate tracheal sound analysis models should be developed.

## 6. Conclusion

The automated analysis of lung and tracheal sounds is of clinical importance. Lung sound and tracheal sound have different acoustic features. Hence, the automated inhalation, exhalation, and CAS detection model trained on lung sounds performed poorly for tracheal sound analysis, and vice versa. However, mixed set training and domain adaptation can improve the performance of models 1) for inhalation and exhalation detection in lung sound analysis and 2) for inhalation, exhalation, and CAS detection in tracheal sound analysis relative to the PCs (lung models trained only by a lung sound and vice versa). In particular, a model derived from mixed set training has great flexibility allowing a user not to select a specific model for lung or tracheal sound analysis, which facilitates the setup of respiratory monitoring in a busy operating room, ward, or clinic.

## Conflict of interest

## Acknowledgements

The tracheal sound collection was contributed by Dr. Ying-Chieh Hsu at Taipei Tzu Chi Hospital, Buddhist Tzu Chi Medical Foundation, Dr. Chang-Fu Su at En Chu Kong Hospital, Dr. Ying-Che Huang at Taipei City Hospital, and Dr. Shou-Zen Fan and Dr. Chung-Chih Shih at National Taiwan University Hospital in Taiwan. The sound collection was sponsored by Raising Children Medical Foundation, Taiwan. The authors thank the employees of Heroic Faith Medical Science for partially contributing to this study. The author would like to thank the National Center for High-performance

## References

[1] Bohadana A, Izbicki G, Kraman SS. Fundamentals of lung auscultation. New England Journal of Medicine. 2014;370:744-51.

[2] Pramono RXA, Bowyer S, Rodriguez-Villegas E. Automatic adventitious respiratory sound analysis: A systematic review. PloS one. 2017;12:e0177926.

[3] Sarkar M, Madabhavi I, Niranjan N, Dogra M. Auscultation of the respiratory system. Annals of thoracic medicine. 2015;10:158.

[4] Ouchi K, Fujiwara S, Sugiyama K. Acoustic method respiratory rate monitoring is useful in patients under intravenous anesthesia. Journal of clinical monitoring and computing. 2017;31:59-65.

[5] Ramsay MA, Usman M, Lagow E, Mendoza M, Untalan E, De Vol E. The accuracy, precision and reliability of measuring ventilatory rate and detecting ventilatory pause by rainbow acoustic monitoring and capnometry. Anesthesia & Analgesia. 2013;117:69-75.

[6] Cathain EO, Gaffey MM. Upper airway obstruction. StatPearls [Internet]: StatPearls Publishing; 2022.

[7] Boriosi JP, Zhao Q, Preston A, Hollman GA. The utility of the pretracheal stethoscope in detecting ventilatory abnormalities during propofol sedation in children. Pediatric Anesthesia. 2019;29:604-10.

[8] Yadollahi A, Giannouli E, Moussavi Z. Sleep apnea monitoring and diagnosis based on pulse oximetery and tracheal sound signals. Medical & biological engineering & computing. 2010;48:1087-97.

[9] Yu L, Ting C-K, Hill BE, Orr JA, Brewer LM, Johnson KB, et al. Using the entropy of tracheal sounds to detect apnea during sedation in healthy nonobese volunteers. Anesthesiology. 2013;118:1341-9.

[10] Liu J, Ai C, Zhang B, Wang Y, Brewer LM, Ting C-K, et al. Tracheal sounds accurately detect apnea in patients recovering from anesthesia. Journal of clinical monitoring and computing. 2019;33:437-44.

[11] Lu X, Azevedo Coste C, Nierat M-C, Renaux S, Similowski T, Guiraud D. Respiratory Monitoring Based on Tracheal Sounds: Continuous Time-Frequency Processing of the Phonospirogram Combined with Phonocardiogram-Derived Respiration. Sensors. 2021;21:99.

[12] Gelb AW, Morriss WW, Johnson W, Merry AF. World Health Organization-World Federation of Societies of Anaesthesiologists (WHO-WFSA) international standards for a safe practice of anesthesia. Canadian Journal of Anesthesia/Journal canadien d'anesthésie. 2018;65:698-708.

[13] Krmpotic K, Rieder MJ, Rosen D. Recommendations for procedural sedation in infants, children, and adolescents. Paediatrics & Child Health. 2021;26:128-.

[14] Association AD. Guidelines for the use of sedation and general anesthesia by dentists. Adopted by the ADA House of Delegates. 2016.

[15] Coté CJ, Wilson S, Pediatrics AAo, Dentistry AAoP. Guidelines for monitoring and management of pediatric patients before, during, and after sedation for diagnostic and therapeutic procedures. Pediatrics. 2019;143.

[16] Hsu F-S, Huang S-R, Huang C-W, Huang C-J, Cheng Y-R, Chen C-C, et al. Benchmarking of eight recurrent neural network variants for breath phase and adventitious sound detection on a self-developed open-access lung sound database-HF_Lung_V1. PLoS One. 2021;16:e0254134.

[17] Kim Y, Hyon Y, Lee S, Woo S-D, Ha T, Chung C. The coming era of a new auscultation system for analyzing respiratory sounds. BMC Pulmonary Medicine. 2022;22:1-11.

[18] Earis J, Cheetham B. Current methods used for computerized respiratory sound analysis. European Respiratory Review. 2000;10:586-90.

[19] Gurung A, Scrafford CG, Tielsch JM, Levine OS, Checkley W. Computerized lung sound analysis as diagnostic aid for the detection of abnormal lung sounds: a systematic review and meta-analysis. Respiratory medicine. 2011;105:1396-403.

[20] Muthusamy PD, Sundaraj K, Abd Manap N. Computerized acoustical techniques for respiratory flow-sound analysis: a systematic review. Artificial Intelligence Review. 2020;53:3501-74.

[21] Rocha BM, Pessoa D, Marques A, Carvalho P, Paiva RP. Automatic classification of adventitious respiratory sounds: A (un) solved problem? Sensors. 2020;21:57.

[22] Hestness J, Narang S, Ardalani N, Diamos G, Jun H, Kianinejad H, et al. Deep learning scaling is predictable, empirically. arXiv preprint arXiv:171200409. 2017.

[23] Sun C, Shrivastava A, Singh S, Gupta A. Revisiting unreasonable effectiveness of data in deep learning era. Proceedings of the IEEE international conference on computer vision2017. p. 843-52.

[24] Hsu F-S, Huang S-R, Huang C-W, Cheng Y-R, Chen C-C, Hsiao J, et al. A Progressively Expanded Database for Automated Lung Sound Analysis: An Update. Applied Sciences. 2022;12:7623.

[25] Weiss K, Khoshgoftaar TM, Wang D. A survey of transfer learning. Journal of Big data. 2016;3:1-40.

[26] Xu W, He J, Shu Y. Transfer Learning and Deep Domain Adaptation. Advances in Deep Learning: IntechOpen; 2020.

[27] Xia T, Han J, Mascolo C. Exploring machine learning for audio-based respiratory condition screening: A concise review of databases, methods, and open issues. Experimental Biology and Medicine. 2022:15353702221115428.

[28] Choi Y, Choi H, Lee H, Lee S, Lee H. Lightweight Skip Connections with Efficient Feature Stacking for Respiratory Sound Classification. IEEE Access. 2022.

[29] Li J, Wang X, Wang X, Qiao S, Zhou Y. Improving The ResNet-based Respiratory Sound Classification Systems With Focal Loss. 2022 IEEE Biomedical Circuits and Systems Conference

(BioCAS): IEEE; 2022. p. 223-7.

[30] Babu N, Kumari J, Mathew J, Satija U, Mondal A. Multiclass Categorisation of Respiratory Sound Signals using Neural Network.  2022 IEEE Biomedical Circuits and Systems Conference (BioCAS): IEEE; 2022. p. 228-32.

[31] Lella KK, Pja A. Automatic diagnosis of COVID-19 disease using deep convolutional neural network with multi-feature channel from respiratory sound data: cough, voice, and breath. Alexandria Engineering Journal. 2022;61:1319-34.

[32] Acharya J, Basu A. Deep neural network for respiratory sound classification in wearable devices enabled by patient specific model tuning. IEEE transactions on biomedical circuits and systems. 2020;14:535-44.

[33] Fraiwan M, Fraiwan L, Alkhodari M, Hassanin O. Recognition of pulmonary diseases from lung sounds using convolutional neural networks and long short-term memory. Journal of Ambient Intelligence and Humanized Computing. 2022;13:4759-71.

[34] Chen H, Yuan X, Pei Z, Li M, Li J. Triple-classification of respiratory sounds using optimized s-transform and deep residual networks. IEEE Access. 2019;7:32845-52.

[35] Grzywalski T, Piecuch M, Szajek M, Bręborowicz A, Hafke-Dys H, Kociński J, et al. Practical implementation of artificial intelligence algorithms in pulmonary auscultation examination. European Journal of Pediatrics. 2019;178:883-90.

[36] Jácome C, Ravn J, Holsbø E, Aviles-Solis JC, Melbye H, Ailo Bongo L. Convolutional neural network for breathing phase detection in lung sounds. Sensors. 2019;19:1798.

[37] Messner E, Fediuk M, Swatek P, Scheidl S, Smolle-Juttner F-M, Olschewski H, et al. Crackle and breathing phase detection in lung sounds with deep bidirectional gated recurrent neural networks. 2018 40th Annual International Conference of the IEEE Engineering in Medicine and Biology Society (EMBC): IEEE; 2018. p. 356-9.

[38] Nakano H, Furukawa T, Tanigawa T. Tracheal sound analysis using a deep neural network to detect sleep apnea. Journal of Clinical Sleep Medicine. 2019;15:1125-33.

[39] Korompili G, Kouvaras M, Kokkalas L, Tatlas NA, Mitilineos S, Potirakis S. Tracheal Sounds, Deep Neural Network, Classification to Distinguish Obstructed from Normal Breathing During Sleep. Forum Acusticum2020. p. 3209-15.

[40] Anesthesiologists ASo. Continuum of depth of sedation: definition of general anesthesia and levels of sedation/analgesia. Approved by ASA House of Delegates on October 13, 1999, and last amended on October 15, 2014. 2014.

[41] Hsu F-S, Huang C-J, Kuo C-Y, Huang S-R, Cheng Y-R, Wang J-H, et al. Development of a respiratory sound labeling software for training a deep learning-based respiratory sound analysis model. International Forum on Medical Imaging in Asia 2021: SPIE; 2021. p. 109-14.

[42] Hsu F-S, Huang S-R, Huang C-W, Cheng Y-R, Chen C-C, Hsiao J, et al. An Update of a Progressively Expanded Database for Automated Lung Sound Analysis. arXiv preprint arXiv:210204062. 2021.

[43] Cohen L. Time-frequency analysis: Prentice Hall PTR Englewood Cliffs, NJ; 1995.

[44] He K, Girshick R, Dollár P. Rethinking imagenet pre-training. Proceedings of the IEEE/CVF International Conference on Computer Vision2019. p. 4918-27.

[45] Goettel N, Herrmann MJ. Breath Sounds: From Basic Science to Clinical Practice. Anesthesia & Analgesia. 2019;128:e42.

[46] Acres JC, Kryger MH. Upper airway obstruction. Chest. 1981;80:207-11.

[47] Hanamoto H, Sugimura M, Morimoto Y, Kudo C, Boku A, Niwa H. Cough reflex under intravenous sedation during dental implant surgery is more frequent during procedures in the maxillary anterior region. Journal of Oral and Maxillofacial Surgery. 2013;71:e158-e63.

[48] Maddox RR, Williams CK, Oglesby H, Butler B, Colclasure B. Clinical experience with patient-controlled analgesia using continuous respiratory monitoring and a smart infusion system. American journal of health-system pharmacy. 2006;63:157-64.

[49] Friesen RH, Alswang M. End-tidal PCO 2 monitoring via nasal cannulae in pediatric patients: accuracy and sources of error. Journal of clinical monitoring. 1996;12:155-9.

[50] Hardman J, Curran J, Mahajan R. End-tidal carbon dioxide measurement and breathing system filters. Anaesthesia. 1997;52:646-8.

[51] Patino M, Redford DT, Quigley TW, Mahmoud M, Kurth CD, Szmuk P. Accuracy of acoustic respiration rate monitoring in pediatric patients. Pediatric Anesthesia. 2013;23:1166-73.

[52] Ahmed I, Aziz E, Newton N. Connection of capnography sampling tube to an intravenous cannula. Anaesthesia. 2005;60:824-5.

[53] Cacho G, Pérez-Calle J, Barbado A, Lledó J, Ojea R, Fernández-Rodríguez C. Capnography is superior to pulse oximetry for the detection of respiratory depression during colonoscopy. Revista espanola de enfermedades digestivas. 2010;102:86.

[54] Lam T, Nagappa M, Wong J, Singh M, Wong D, Chung F. Continuous pulse oximetry and capnography monitoring for postoperative respiratory depression and adverse events: a systematic review and meta-analysis. Anesthesia & Analgesia. 2017;125:2019-29.

# Supplementary

The performance indexes of segment detection are presented in the supplementary Table S1.

**Table S1. Performance of segment detection of the trained models.**

| Controls | Training Database/Strategy | Testing Database | Segment Detection | | | | | |
|---|---|---|---|---|---|---|---|---|
| | | | ACC | PPV | SEN | SPE | F1-Score | AUC |
| Inhalation | | | | | | | | |
| PC | Lung_V2 | Lung_V2 | 0.931±0.003 | 0.844±0.011 | 0.799±0.012 | 0.963±0.003 | 0.821±0.008 | 0.973±0.002 |
| NC | Tracheal_V1 | Lung_V2 | 0.892±0.004††† | 0.764±0.011††† | 0.661±0.017††† | 0.949±0.003††† | 0.709±0.012††† | 0.934±0.004††† |
| | Lung_V2+Tracheal_V1 | Lung_V2 | **0.933±0.002*** | **0.849±0.008** | **0.805±0.010** | **0.965±0.003** | **0.826±0.004*** | **0.975±0.001***** |
| | Tracheal_V1→Lung_V2 | Lung_V2 | 0.931±0.003 | 0.847±0.009 | 0.801±0.010 | 0.964±0.003 | 0.823±0.007 | 0.974±0.001 |
| NC | Lung_V2 | Tracheal_V1 | 0.888±0.007††† | 0.778±0.016††† | 0.616±0.027††† | 0.956±0.003††† | 0.687±0.022††† | 0.916±0.010 |
| PC | Tracheal_V1 | Tracheal_V1 | 0.931±0.002 | 0.855±0.007 | 0.785±0.011 | **0.967±0.002** | 0.818±0.006 | 0.971±0.002 |
| | Lung_V2+Tracheal_V1 | Tracheal_V1 | 0.930±0.002 | 0.854±0.009 | 0.783±0.013 | 0.967±0.003 | 0.817±0.007 | 0.970±0.002 |
| | Lung_V2→Tracheal_V1 | Tracheal_V1 | **0.934±0.002**** | **0.858±0.005** | **0.800±0.012**** | 0.967±0.002 | **0.828±0.006**** | **0.974±0.001**** |
| Exhalation | | | | | | | | |
| PC | Lung_V2 | Lung_V2 | 0.924±0.003 | 0.783±0.011 | 0.644±0.019 | 0.970±0.002 | 0.706±0.012 | 0.954±0.002 |
| NC | Tracheal_V1 | Lung_V2 | 0.887±0.003††† | 0.665±0.016††† | 0.416±0.029††† | 0.965±0.004††† | 0.511±0.023††† | 0.881±0.008††† |
| | Lung_V2+Tracheal_V1 | Lung_V2 | **0.928±0.001**** | **0.796±0.009**** | **0.667±0.018**** | **0.972±0.002** | **0.726±0.008**** | **0.958±0.002**** |
| | Tracheal_V1→Lung_V2 | Lung_V2 | 0.926±0.001* | 0.789±0.009 | 0.656±0.011* | 0.971±0.002 | 0.716±0.005* | 0.957±0.001*** |
| NC | Lung_V2 | Tracheal_V1 | 0.873±0.008††† | 0.757±0.023††† | 0.445±0.050††† | 0.968±0.005 | 0.559±0.043††† | 0.865±0.014††† |
| PC | Tracheal_V1 | Tracheal_V1 | 0.928±0.002 | 0.840±0.006 | 0.751±0.010 | 0.968±0.002 | 0.793±0.005 | **0.963±0.002** |
| | Lung_V2+Tracheal_V1 | Tracheal_V1 | 0.929±0.001 | 0.846±0.006* | 0.748±0.009 | 0.970±0.002** | 0.794±0.004 | 0.962±0.002 |
| | Lung_V2→Tracheal_V1 | Tracheal_V1 | **0.931±0.001**** | **0.852±0.006**** | **0.752±0.007** | **0.971±0.002**** | **0.799±0.004**** | 0.963±0.002 |

| Controls | Training Database/Strategy | Testing Database | Segment Detection | | | | | |
|---|---|---|---|---|---|---|---|---|
| | | | ACC | PPV | SEN | SPE | F1-Score | AUC |
| Continuous adventitious sound | | | | | | | | |
| PC | Lung_V2 | Lung_V2 | **0.884±0.003** | **0.675±0.013** | **0.495±0.028** | 0.956±0.004 | **0.571±0.018** | **0.918±0.005** |
| NC | Tracheal_V1 | Lung_V2 | 0.861±0.002 | 0.590±0.008 | 0.342±0.025 | 0.956±0.004 | 0.432±0.020 | 0.854±0.004 |
| | Lung_V2+Tracheal_V1 | Lung_V2 | 0.881±0.003 | 0.671±0.012 | 0.461±0.028 | **0.958±0.004** | 0.546±0.020 | 0.914±0.005 |
| | Tracheal_V1→Lung_V2 | Lung_V2 | 0.883±0.003 | 0.670±0.010 | 0.493±0.044 | 0.955±0.005 | 0.567±0.029 | 0.917±0.005 |
| NC | Lung_V2 | Tracheal_V1 | 0.881±0.011 | 0.721±0.053 | 0.413±0.069 | 0.970±0.007 | 0.522±0.066 | 0.886±0.012 |
| PC | Tracheal_V1 | Tracheal_V1 | 0.934±0.005 | 0.859±0.029 | 0.702±0.071 | 0.977±0.008 | 0.770±0.029 | 0.972±0.003 |
| | Lung_V2+Tracheal_V1 | Tracheal_V1 | 0.946±0.006 | **0.870±0.021** | 0.777±0.029 | **0.978±0.004** | 0.821±0.021 | 0.977±0.004 |
| | Lung_V2→Tracheal_V1 | Tracheal_V1 | **0.949±0.002** | 0.860±0.014 | **0.814±0.018** | 0.975±0.003 | **0.836±0.008** | **0.981±0.002** |

PC: positive control, and NC: negative control. Bold values indicate better performance. Lung_V2+Tracheal_V1 denotes mixed set training. Tracheal_V1→Lung_V2 denotes pretrained tracheal sound model that was fine-tuned with the lung sound data. Lung_V2→Tracheal_V1 denotes the pretrained lung sound model that was fine-tuned with the tracheal sound data. All values are presented as mean ± standard deviation.

†, ††, and ††† represent significantly worse performance ($p < 0.05$, $p < 0.01$, and $p < 0.001$, respectively) compared with PC.

*, **, and *** represent significantly better performance ($p < 0.05$, $p < 0.01$, and $p < 0.001$, respectively) compared with PC.